\documentclass[conference]{IEEEtran}
\IEEEoverridecommandlockouts
\usepackage{cite}
\usepackage{amsmath,amssymb,amsfonts}
\usepackage{graphicx}
\usepackage{textcomp}
\usepackage{xcolor}
\usepackage{algorithm}
\usepackage{algorithmicx}
\usepackage{algpseudocode}
\usepackage{subcaption}
\usepackage{booktabs}

\usepackage[a4paper, total={184mm,239mm}]{geometry}
\def\BibTeX{{\rm B\kern-.05em{\sc i\kern-.025em b}\kern-.08em
    T\kern-.1667em\lower.7ex\hbox{E}\kern-.125emX}}
\begin{document}

\title{SpecHD: Hyperdimensional Computing Framework for FPGA-based Mass Spectrometry Clustering\\
}

\author{
    \IEEEauthorblockN{
        Sumukh Pinge\IEEEauthorrefmark{1},
        Weihong Xu\IEEEauthorrefmark{1},
        Jaeyoung Kang\IEEEauthorrefmark{1},
        Tianqi Zhang\IEEEauthorrefmark{1},
        Niema Moshiri\IEEEauthorrefmark{1},
        Wout Bittremieux\IEEEauthorrefmark{2},
        Tajana Rosing\IEEEauthorrefmark{1}
    }
    \IEEEauthorblockA{\IEEEauthorrefmark{1}University of California, San Diego, US}
    \IEEEauthorblockA{\IEEEauthorrefmark{2}University of Antwerp, BE}
}

\maketitle

\begin{abstract}
Mass spectrometry-based proteomics is a key enabler for personalized healthcare, providing a deep dive into the complex protein compositions of biological systems. This technology has vast applications in biotechnology and biomedicine but faces significant computational bottlenecks. Current methodologies often require multiple hours or even days to process extensive datasets, particularly in the domain of spectral clustering. To tackle these inefficiencies, we introduce Spec-HD, a hyperdimensional computing framework supplemented by an FPGA-accelerated architecture. Utilizing streamlined binary operations in a hyperdimensional computational environment, Spec-HD capitalizes on the low-latency and parallel capabilities of FPGAs. This approach markedly improves clustering speed and efficiency, serving as a catalyst for real-time, high-throughput data analysis in future healthcare applications. Our evaluations demonstrate that Spec-HD not only maintains but often surpasses existing clustering quality metrics while drastically cutting computational time. Specifically, it can cluster a large-scale human proteome dataset—comprising 25 million MS/MS spectra and 131 GB of MS data—in just 5 minutes. With energy efficiency exceeding 31× and a speedup factor that spans a range of 6× to 54× over existing state-of-the-art solutions, Spec-HD emerges as a promising solution for the rapid analysis of mass spectrometry data with great implications for personalized healthcare.

\end{abstract}

\begin{IEEEkeywords}
Mass spectrometry, Proteomics, Spectral clustering, Hyperdimensional computing.
\end{IEEEkeywords}

\section{Introduction}

Mass spectrometry (MS) is a cornerstone technique in proteomics research, enabling the detailed examination of protein compositions in various biological samples. With the progressive evolution of MS technologies, there has been a corresponding surge in data volumes—in the order of terabytes per month. This voluminous data offers unparalleled insights into the intricate world of proteins. Further accentuating this trend, MS data repositories like MassIVE are expanding at an unprecedented rate, amassing over 500TB of data as of September 2023. These vast datasets harbor the potential to revolutionize personalized medicine by facilitating the identification of patient-specific biomarkers\cite{ciocan-cartita2019relevance}. However, the sheer volume of spectra generated in typical MS experiments poses significant computational challenges, especially in tasks like spectral clustering. 

Clustering in MS/MS spectra groups similar data, producing representative consensus spectra. This not only reduces redundancy but also accelerates the database searching—a significant bottleneck in proteomic data analysis. By fine-tuning the data, spectral clustering substantially refines the peptide identification process, enhancing both its speed and precision. In fact, this optimization can reduce the runtime of peptide identification by up to 50\%\cite{mscrush}. The benefits derived from clustering become particularly salient in personalized healthcare settings. Here, expedited data analysis can critically influence the timeliness and quality of patient care. However, despite its advantages, spectral clustering is often underutilized due to its time-consuming nature and the limitations of existing tools. MS clustering, unlike traditional methods, grapples with high-dimensional, noisy spectral data. Each MS spectrum contains hundreds to thousands of intensity values, necessitating specialized algorithms sensitive to such data intricacies. While there are FPGA-based solutions for clustering, they may not be explicitly optimized for MS's unique challenges. In response, we introduce SpecHD, rooted in hyperdimensional computing. Primarily for FPGA-accelerated mass spectrometry clustering, SpecHD capitalizes on FPGA's parallel processing and low latency. Beyond clustering, SpecHD adeptly handles the computational demands of data preprocessing, a pivotal step often sidelined in the MS data workflow. The following are key highlights of SpecHD's contributions and advantages in the realm of MS data processing:

\begin{enumerate}
    \item SpecHD is among the first to implement the linkage agnostic NN-Chain Hierarchical Agglomerative clustering algorithm using FPGAs. This strategic alignment boosts MS clustering speed by 6 -- 54× and achieves an energy efficiency 31× greater than current benchmarks.
    
    \item With SpecHD, we synergistically blend near-storage MS preprocessing with FPGA capabilities. Guided by design space exploration, this combination yields notable advancements in both hardware efficiency and energy conservation, facilitating seamless data exchanges between the FPGA and NVMe storage.
    
    \item Unlike conventional end-to-end tools, SpecHD introduces a refined approach. By storing spectral data in the hyperdimensional space, we achieve significant data compression. Our analysis also highlights SpecHD's superiority in clustering quality and enhanced database search results compared to its MS clustering counterparts.
\end{enumerate}

\section{Background}

\subsection{Mass-Spectrometry Pipeline}
\begin{figure}[h]
\centering
\includegraphics[width=0.4\textwidth]{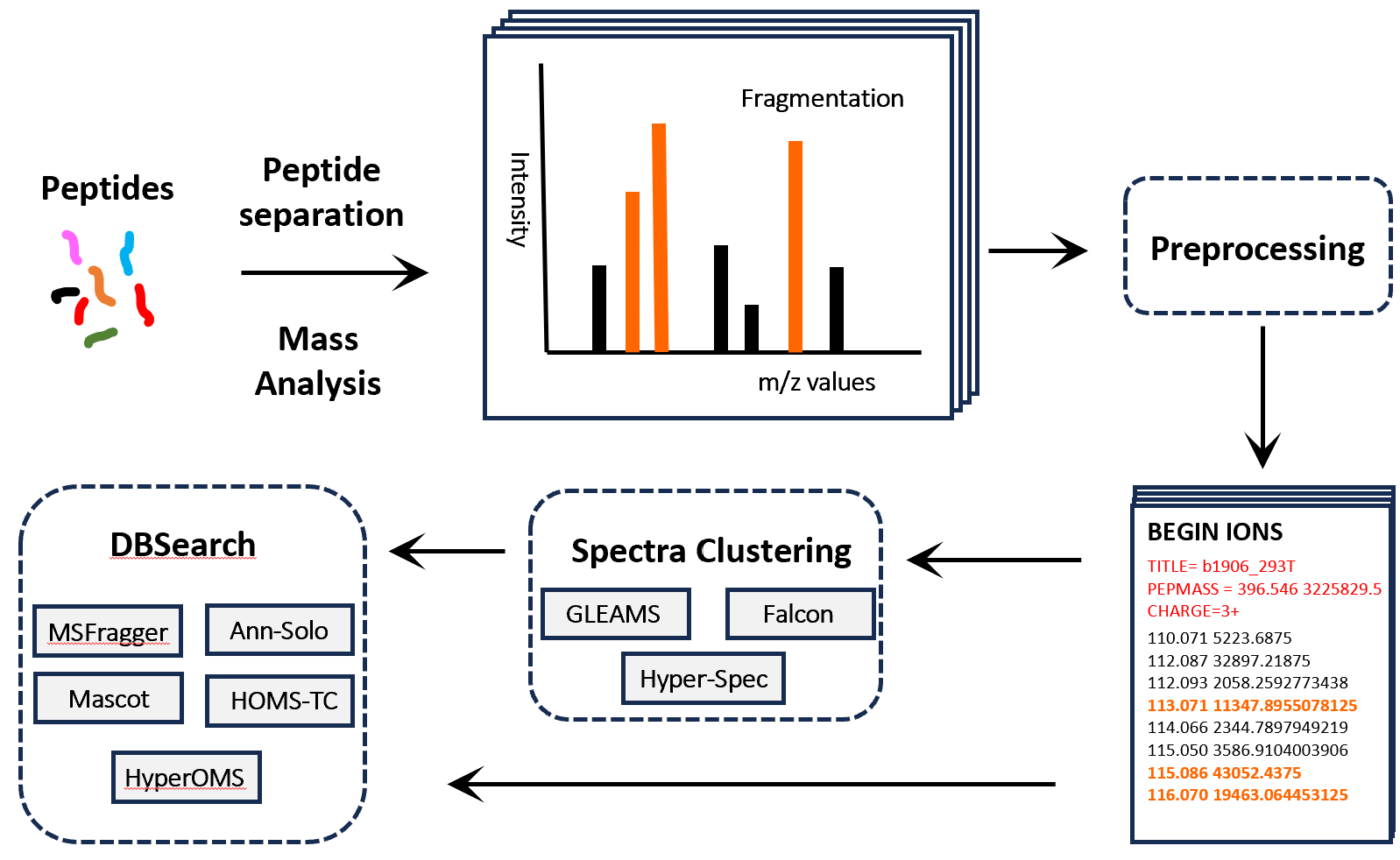}
\caption{MS data-analysis pipeline}
\label{fig:pipeline}
\end{figure}

Mass Spectrometry operates as an intricate information processing pipeline, transforming a biological sample into a digital spectral representation (Fig.~\ref{fig:pipeline}). Initiated by the ionization of the specimen, charged ions are passed through a mass analyzer and sorted by their mass-to-charge (m/z) ratios, providing a quantitative and qualitative snapshot of these values. The resulting data is then converted into structured digital formats, namely mzML, mzXML, Mascot Generic File (mgf), and MS2 files. These files plot m/z ratios against ion intensities, converting the spectral peaks into computational vectors. This data is primed for database searching, a task similar to pattern matching in large datasets. This crucial step matches the observed spectra with known peptide sequences, both identifying proteins in the original sample and bridging raw data to biologically relevant insights. By translating spectral peaks and intensities into potential protein identifications, researchers can uncover crucial insights. Within the scope of personalized healthcare, the precise identification of these proteins plays a key role, assisting in the detection of disease-specific biomarkers and shaping individualized treatment strategies.

\subsection{Related Works}

MS clustering tools face the dual challenge of maximizing clustering quality while minimizing computational time.  Various methods address these issues in unique ways. For instance, MaRaCluster\cite{the2016maracluster} employs optimized distance metrics to improve clustering quality. Falcon\cite{bittremieux2021largescale} uses hashing techniques to reduce dimensionality and employs approximate nearest neighbor algorithms to expedite computations. Similarly, msCRUSH\cite{mscrush} employs locality-sensitive hashing to avoid unnecessary pairwise comparisons between spectra. GLEAMS\cite{bittremieux2022learned} takes a different approach by using a deep neural network trained in a supervised manner, embedding spectra in a low-dimensional space for optimized clustering. ClusterSheep enhances computational speed by leveraging GPUs and specialized function kernels. However, these approaches often make trade-offs between quality and speed.

 Bridging the gap between previous methods, HyperSpec \cite{xu2023hyperspec} has emerged as a promising tool, utilizing hyperdimensional computing (HDC) to improve both speed and quality. HDC offers a way to encode spectra into high-dimensional vectors, known as hypervectors (HVs), providing compact, binary vector representations with minimal loss of information. This approach was implemented on GPUs, achieving state-of-the-art results in speed with comparable clustering quality. However, a significant concern with HyperSpec and other GPU-based solutions is when datasets surpass the GPU's onboard memory capacity. This constraint inhibits efficient data processing and frequently necessitates data transfers between the GPU and the main memory, leading to performance overheads. Additionally, the increased power consumption of GPUs, particularly at peak operations, adds to operational costs \cite{pattnaik2016scheduling}. Beyond these challenges, HyperSpec is also dependent on general-purpose libraries, offering two flavours of clustering algorithms: DBSCAN via the cuML library and Hierarchical Agglomerative Clustering (HAC) using the fastcluster library, targeting both GPU and CPU platforms, respectively.

 A critical bottleneck in MS clustering tools is the spectra loading and preprocessing step, accounting for an average of 82\% of the total execution time, as corroborated by prior work\cite{xu2022nearstorage}. The escalating growth MS data poses a substantial challenge to current clustering solutions, methods reliant on CPU and GPU architectures, making repository-scale clustering increasingly impractical. Here, we introduce SpecHD, an FPGA-optimized framework for HDC in MS clustering. Distinct from HyperSpec's GPU architecture, SpecHD emphasizes near-storage computing to counteract data transfer bottlenecks associated with GPUs allowing for efficient spectral clustering without sacrificing quality. SpecHD, therefore, aims to address the current gaps in mass spectrometry clustering by combining the best of both worlds: high-quality clustering and computational efficiency.

\subsection{Nearest Neighbor Chain Algorithm}

HyperSpec, which utilizes DBSCAN via the cuML library and Hierarchical Agglomerative Clustering (HAC) through the FastCluster library, has demonstrated effectiveness on both CPU and GPU platforms. While HyperSpec represents a significant advancement and offers a range of capabilities, its reliance on general-purpose libraries and computing architectures may not fully leverage the unique advantages of FPGA architectures, particularly in parallel processing and real-time capabilities, in a rapidly expanding data landscape. Conventional hierarchical clustering algorithms, particularly the classic HAC methods, face computational bottlenecks due to their \(O(n^3)\) time complexity. These algorithms require full matrix updates to calculate pairwise distances between all data points and to identify the minimum distance among all pairs.

In contrast, the Nearest Neighbor Chain (NN-Chain) Algorithm \cite{murtagh2011methods} in SpecHD is tailored for FPGA architectures. The algorithm starts by calculating pairwise distances, akin to traditional methods but streamlines the subsequent computational steps. The NN-Chain Algorithm constructs a `chain' of closest points and evaluates this chain to identify Reciprocal Nearest Neighbors (RNN). Upon identifying an RNN pair, the clusters are merged, and the distance matrix is updated more efficiently, avoiding the need for a full matrix update. This targeted approach minimizes redundant calculations and makes NN-Chain well-suited for large-scale, data-intensive tasks without compromising clustering quality.

 \begin{figure}[h]
\centering
\includegraphics[width=0.45\textwidth]{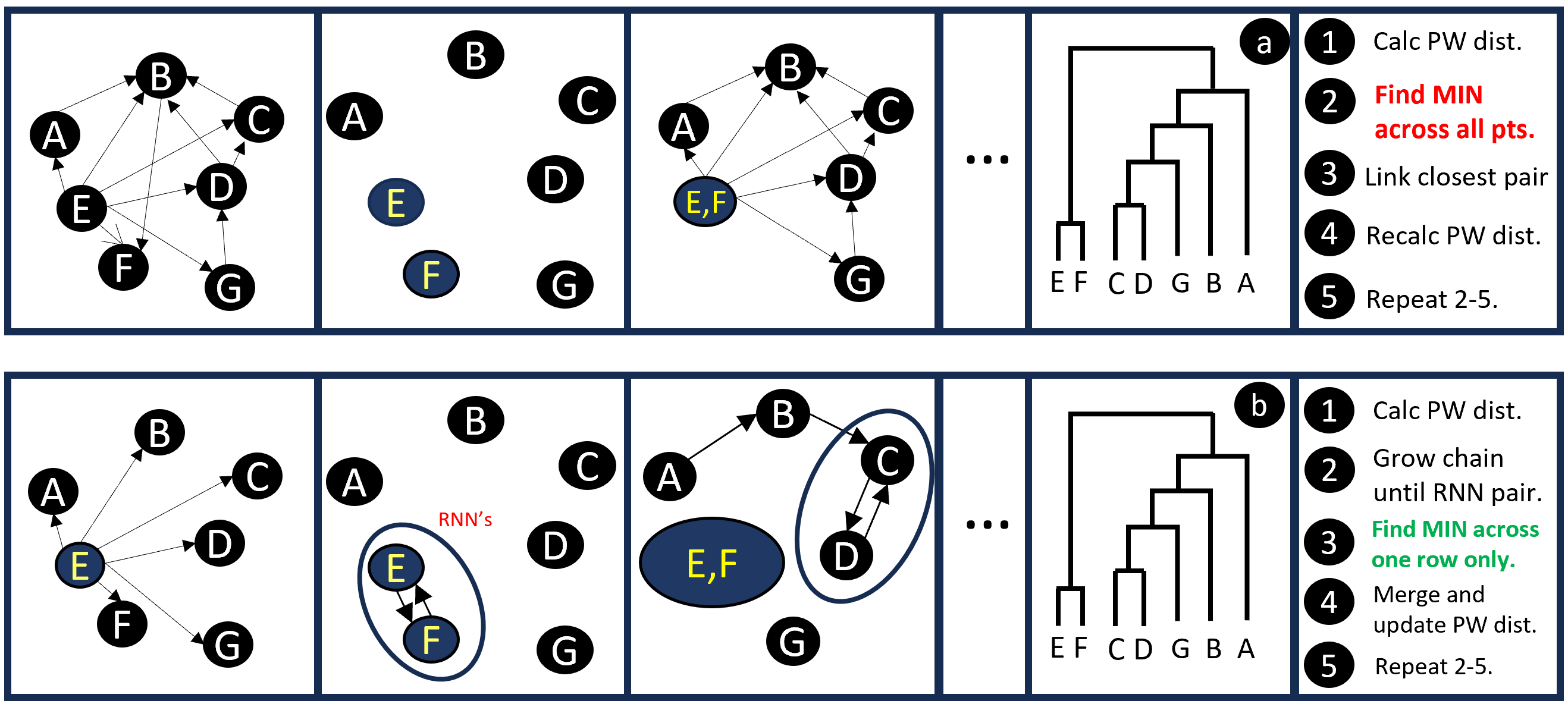}
\caption{Comparison between Naive and NN-chain HAC}
\label{fig:NNCHAIN}
\end{figure}

\section{Methodology and Flow}

 \subsection{Proposed Preprocessing Module}

In the vast landscape of mass spectrometry (MS) tools, certain modules like the Spectra Filter, Top-k Selector, and Scale and Normalization emerge as standard features in MS preprocessing. However, leveraging specialized near-storage frameworks such as MSAS \cite{xu2022nearstorage} can introduce significant performance enhancements.  Our focus is the synergistic integration of MSAS with peer-to-peer (P2P) capabilities, prominently featured in platforms like the U280 Xilinx Alveo. Enabling P2P allows for direct data exchanges between the FPGA and NVMe storage, eliminating intermediary host memory interactions and reducing bandwidth constraints. This capability extends to other Alveo PCIe platforms, provided the host system supports the required large physical address space. The MSAS accelerator, implemented using CMOS technology, is integrated into the same die as the SSD’s embedded cores. It connects to the global on-chip bus and fetches data directly from NAND flashes, achieving peak bandwidth equivalent to external SSDs. Here, the Spectra Filter module stands out by efficiently filtering out peaks related to the precursor ion or with intensities less than 1\% of the base peak, preparing the ground for the Top-k Selector, which employs a streamlined Bitonic sorting algorithm. This design minimizes data movement overhead, largely by filtering out redundant spectral data during the post-preprocessing stage. This sets the stage for optimized, large-scale MS data handling and paves the way for significant improvements in the efficiency of the preprocessing phase. Evaluations and advantages of this approach will be elaborated upon in Section IV.

 \begin{figure}[!h]
\centering
\includegraphics[width=0.49\textwidth]{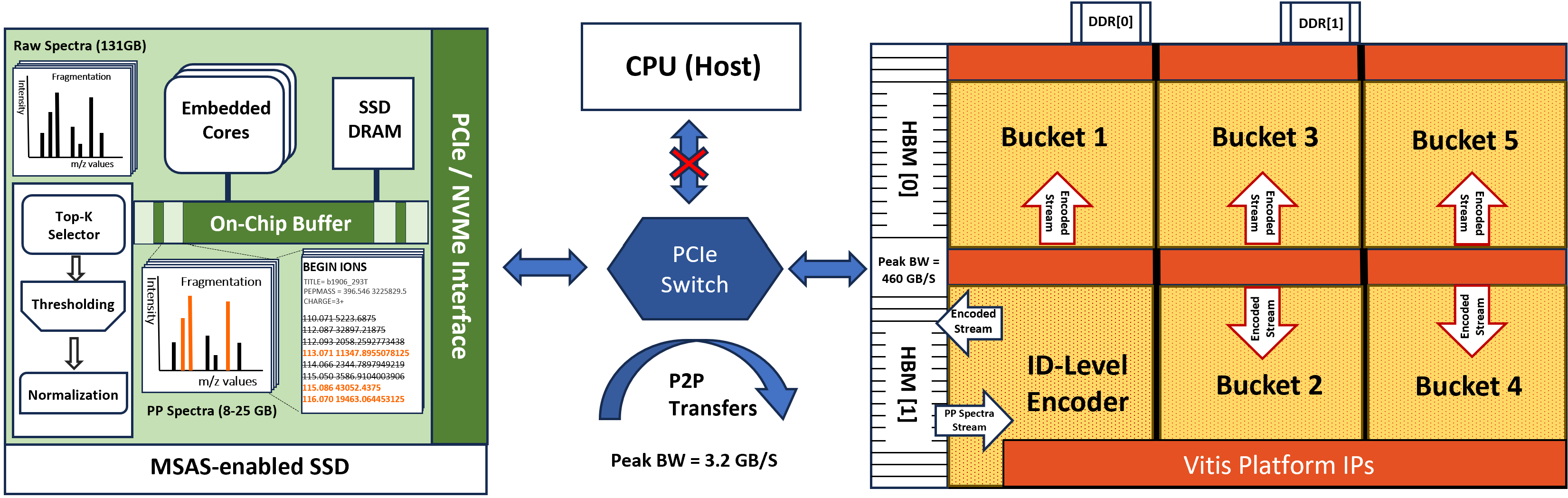}
\caption{Top-level dataflow and kernel arrangement}
\label{fig:toplevel}
\end{figure}

In the context of large-scale datasets, naive approaches to pairwise spectrum comparisons can rapidly escalate into computational bottlenecks, especially given the constraints of on-chip memory. To mitigate this, we adopt a data organization strategy based on precursor \(m/z\) sorting. To manage the computational complexity, we partition the dataset \cite{bittremieux2021largescale} into smaller, discrete `buckets'  calculated as:

\begin{equation}
\text{bucket}_i = \left\lfloor \frac{(m/z_i - 1.00794) \times C_i}{\text{resolution}} \right\rfloor
\end{equation}

Here, \(C_i\) represents the charge state of the \(i^{th}\) spectrum and 1.00794 is the mass of the charge. The term 'resolution' is used to describe the granularity of the mass spectrometer's measurements, and this value can range from 0.05 to 1. This bucketing strategy is particularly beneficial when applied to high-resolution MS spectrometers, as it facilitates to optimize both computational resources and data quality.

 \subsection{ID-Level Encoder}

In the preprocessing phase, the dataset is broken down into smaller, manageable `buckets' of spectra. Each of these spectra consists of two vectors of size \texttt{peak\_count}: one for mass-to-charge ratios (m/z) and another for intensities. Our aim is to efficiently encode these spectra into a single high-dimensional vector of size \texttt{Dhv} through an ID-Level encoding scheme \cite{imani2017voicehd}.

To achieve this, both the m/z values and intensity values are quantized. Pre-allocated vectors from high-dimensional memory spaces, denoted as \texttt{ID[0,f]} for m/z and \texttt{L[0,q]} for intensity, and each of size \texttt{Dhv}, are utilized for this purpose. For each pair of m/z and intensity values, bitwise XOR operations are performed on the corresponding vectors from \texttt{ID} and \texttt{L}. The results are successively accumulated into a single vector until all \texttt{peak\_count} pairs have been processed. A pointwise majority function is then applied to this aggregated vector, culminating in a refined binarized spectrum hypervector: 

\begin{equation}
\text{spectra}_i = \sum_{(i, j)} (I_i \oplus L_j)
\end{equation}

To accelerate the encoding process, the module employs several hardware-level optimizations designed for FPGA platforms. Specifically, data partitioning directives are applied to the \texttt{ID} and \texttt{Level} memory arrays via HLS pragmas. This allows for simultaneous multiple accesses to these arrays, thereby facilitating loop unrolling within the \texttt{hd\_encoding} function. In turn, this minimizes the initiation interval, leading to a significant boost in data throughput. Loop unrolling is further optimized through HLS pragmas, ensuring parallel processing across \texttt{peak\_count}. The resultant high-dimensional vectors are stored in High Bandwidth Memory (HBM), optimizing both memory access and data retrieval speeds for further processing of clustering kernels.

 \begin{figure}[!h]
\centering
\includegraphics[width=0.43\textwidth]{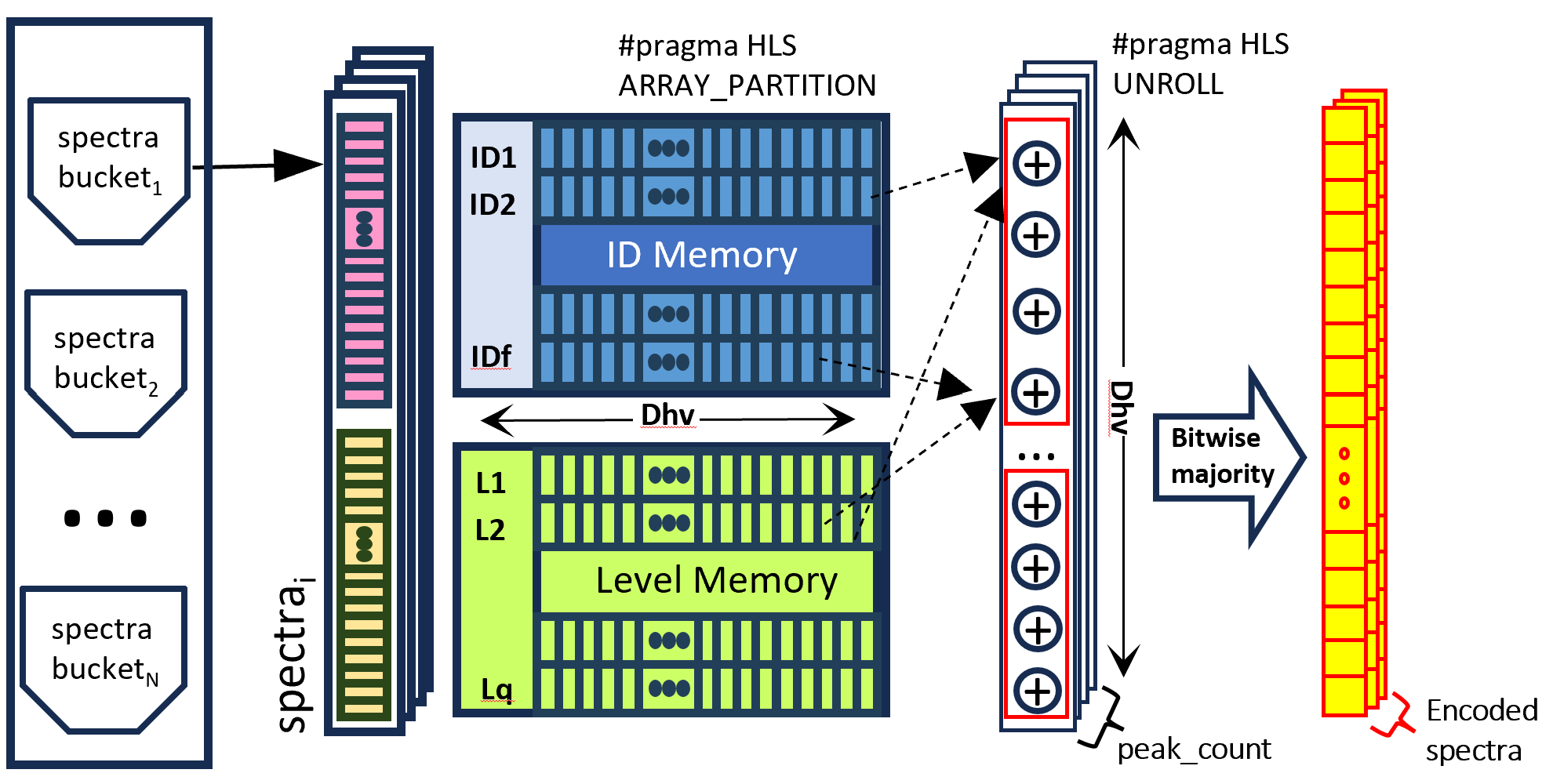}
\caption{ID-Level Encoding}
\label{fig:encoder}
\end{figure}

\subsection{Kernel-Level Acceleration for NN-Chain HAC:}

 Our architecture capitalizes on the robust and adaptable nature of the NN-Chain Algorithm for hierarchical agglomerative clustering (HAC), exploiting the FPGA's inherent parallel processing capabilities. The HLS-optimized kernel function, \texttt{agglomerative\_ccl\_kernel}, functions as the computational core, parameterized to operate on diverse data structures. 

 \begin{figure}[!h]
\centering
\includegraphics[width=0.47\textwidth]{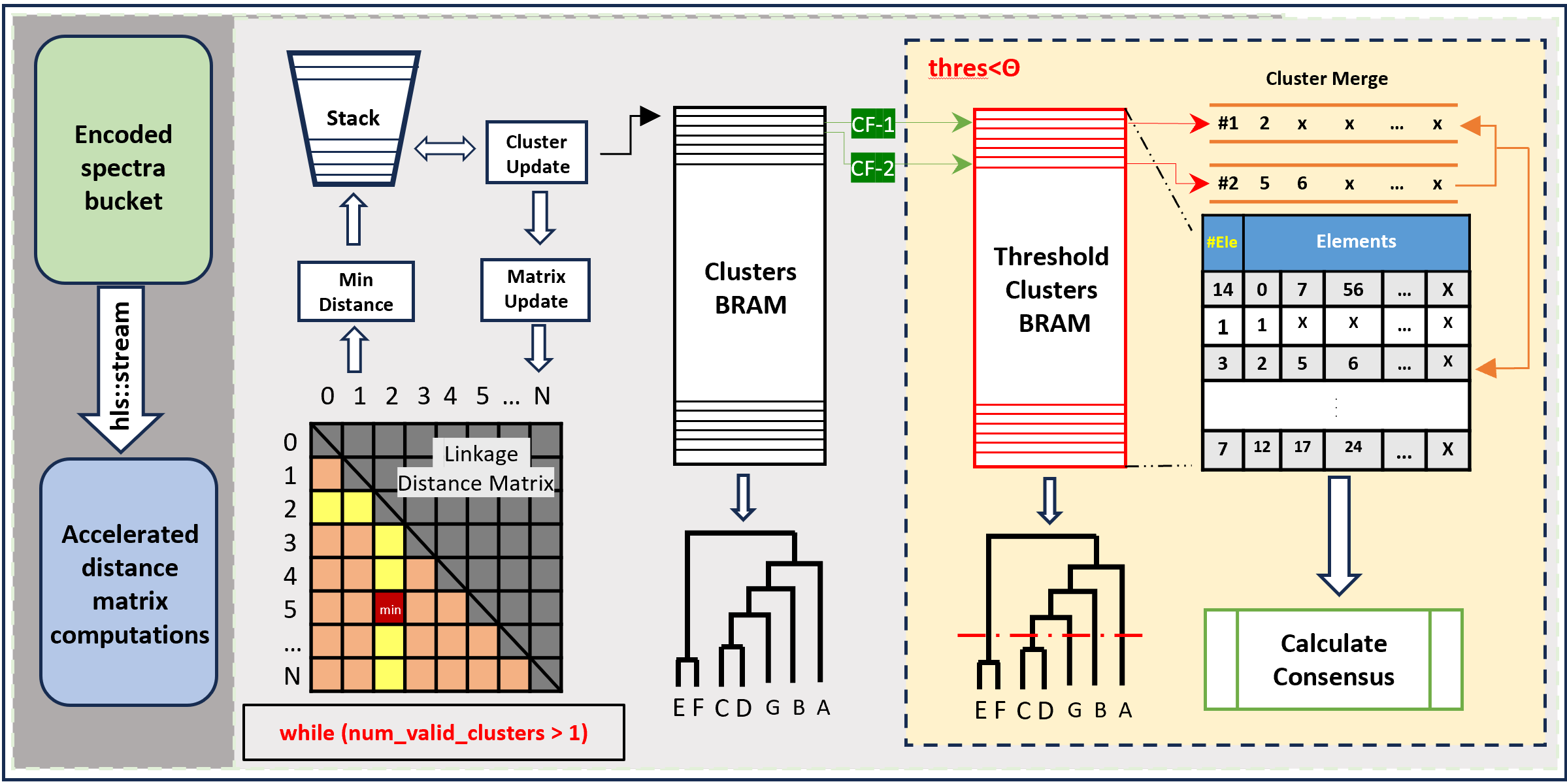}
\caption{FPGA-accelerated NN-Chain HAC architecure}
\label{fig:new_nnchain}
\end{figure}

\vspace*{-2mm}

\textbf{Optimized Distance Matrix Computation:} The architecture incorporates specialized modules, including a fast unrolled XOR and an efficient population count (\textit{popcount}) module, both parameterized for (\texttt{\text{Dhv}}) bits of dimensionality. A dataflow approach is employed, facilitating task-level parallelism by enabling concurrent execution of both reading the encoded spectra and calculating distances, thereby boosting the efficiency of spectra processing and accelerating the computation of the distance matrix. To conserve storage resources, only the lower triangular part of the distance matrix is retained, capitalizing on its symmetry. Furthermore, the use of 16-bit fixed-point arithmetic results in a significant reduction in memory footprint while maintaining computational accuracy.

\textbf{NN-Chain Algorithm:} The process begins by selecting an arbitrary point and calculating its minimum distance to all other points based on a linkage distance matrix. Both the selected point and its nearest neighbor are initially added to a stack. As the algorithm iterates, new elements are added to this stack based on the smallest distance criterion until a Reciprocal Nearest Neighbor (RNN) is identified. Specifically, if the last index in the stack matches the index of the current minimum distance, the algorithm proceeds to cluster merging and updates the distance matrix.

The algorithm manages two separate sets of clusters. One set, stored in Cluster BRAM, is subject to the exhaustive tree traversal, as local chains cannot always be guaranteed to fall under a predefined distance threshold. Another set only merges if the inter-cluster distance is below this threshold. Each cluster is comprised of three key components: the number of elements, the elements themselves, and a correction factor for adaptive adjustments. Upon identifying an RNN, clusters are merged based on their indices, with the second cluster folding into the first. This merging operation updates both the elements and their total count within the cluster. A deleted cluster is effectively removed from future traversals, and its position is replaced by the next cluster in the array. Correction factors are used to dynamically synchronize these cluster updates. Following each cluster merge, the distance matrix is updated based on the chosen linkage criteria. Our architecture is flexible and supports various linkage criteria, including Ward, single linkage, and complete linkage. In our specific implementation, we have found that complete linkage provides the most reliable results. In the concluding steps, the algorithm calculates a consensus cluster by evaluating the lowest average minimum distance to all other spectra within that cluster, based on the original distance matrix. Various optimization techniques, such as memory partitioning and pipelining, are deployed to maximize computational efficiency and throughput. These features make our NN-Chain architecture a robust and adaptable solution for FPGA platforms.

\section{Results}

In our experimental setup, we utilized the Xilinx Alveo U280 Data Center Accelerator Card, featuring an HBM2 total capacity of 8GB and a bandwidth of 460GB/s. For comparative evaluation, our benchmarking setup includes a server with a 12-core CPU, 128GB DDR4 memory, and a 2TB NVMe solid-state drive. An NVIDIA GeForce RTX 3090 GPU with 24GB RAM also formed part of this comparison. We performed extensive design space exploration for SSD-level MSAS accelerator in our preprocessing phase,  targeting both speed and energy optimization. For further evaluations, the preprocessed dataset was channeled from our NVMe SSD storage directly to the HBM using the P2P transfer method, with evaluations employing a single encoder and 5 clustering kernels, mirroring the data flow intrinsic to SpecHD’s architecture as depicted in Fig.~\ref{fig:toplevel}.

\subsection{Pre-processing Results}

To quantify the efficiency of our preprocessing module, we evaluated it across five different datasets, as shown in Table~\ref{tab:preprocess_results}. The hardware setup and configurations are based on an Intel SSD DC P4500, and we have emulated the setup as described in the referenced literature\cite{xu2022nearstorage}. Energy consumption estimates are derived from combining SSD simulation data with the SSD power model referenced \cite{jung2015nandflashsim}.

\begin{table}[h]
    \centering
    \caption{Preprocessing Performance Metrics}
    \label{tab:preprocess_results}
    \scriptsize
    \setlength{\tabcolsep}{6pt}  
    \begin{tabular}{|l|l|l|l|l|l|l|}
    \hline
    Sample Type & PRIDE ID & \#Spectra & Size & PP Time(s) & Energy(J) \\
    \hline
    Kidney cell & PXD001468 & 1.1M & 5.6 GB & 1.79 & 17.38 \\
    \hline
    Kidney cell & PXD001197 & 1.1M & 25 GB & 8.22 & 77.27 \\
    \hline
    HeLa proteins & PXD003258 & 4.1M & 54 GB & 18.44 & 166.53 \\
    \hline
    HEK293 cell & PXD001511 & 4.2M & 87 GB & 28.53 & 268.22 \\
    \hline
    Human proteome  & PXD000561 & 21.1M & 131 GB & 43.38 & 382.62 \\
    \hline
    \end{tabular}
\end{table}

\vspace*{-2mm}

 \subsection{Optimizing Linkage and Data Compression}

\begin{figure}[htbp]
    \centering
    \begin{subfigure}{0.45\linewidth}
        \centering
        \includegraphics[width=1\linewidth]{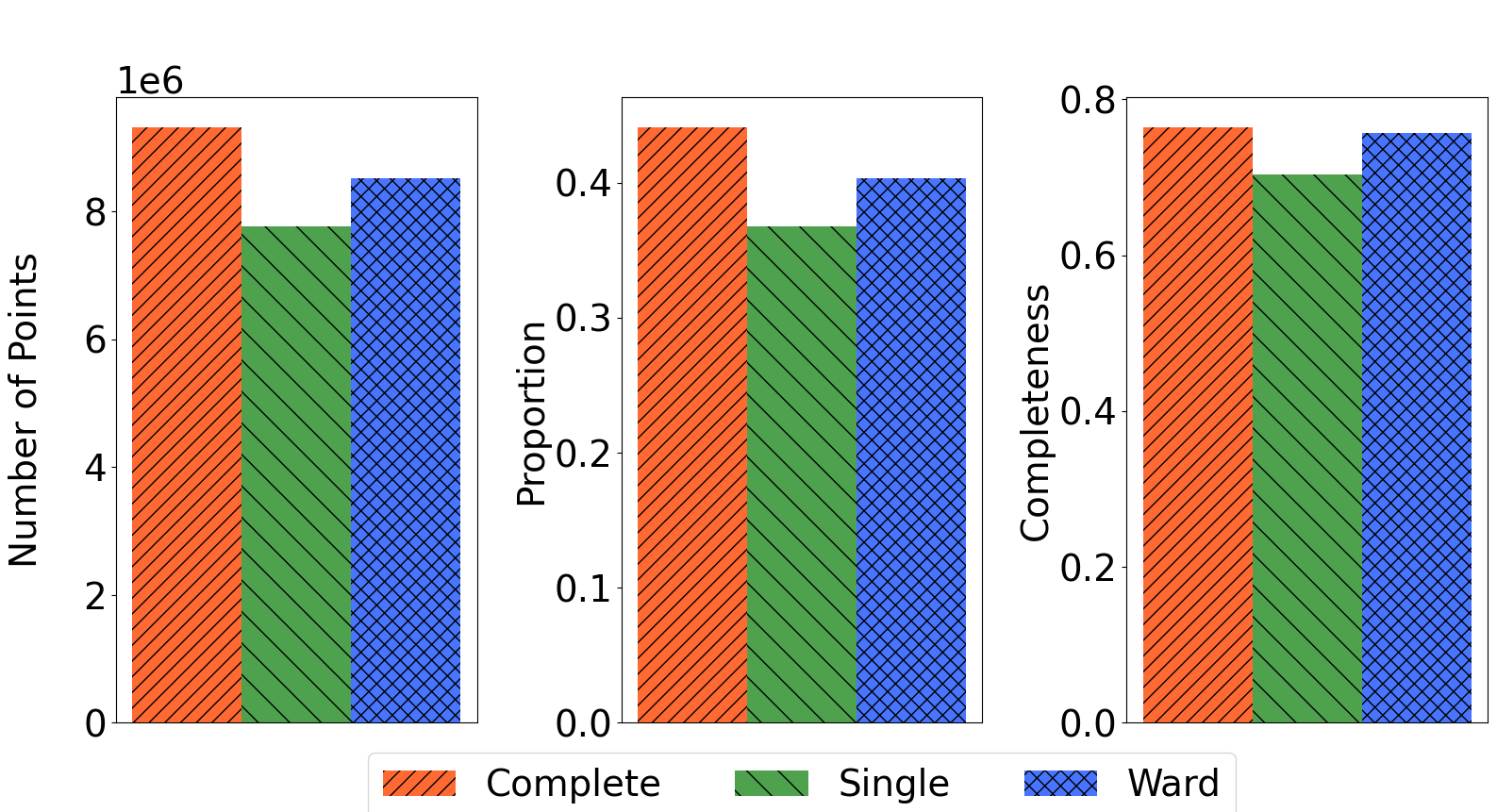}
        \caption{Linkage Comparison}
        \label{fig:Comparison}
    \end{subfigure}
    \hspace{0.01\linewidth} 
    \begin{subfigure}{0.45\linewidth}
        \centering
        \includegraphics[width=0.95\linewidth]{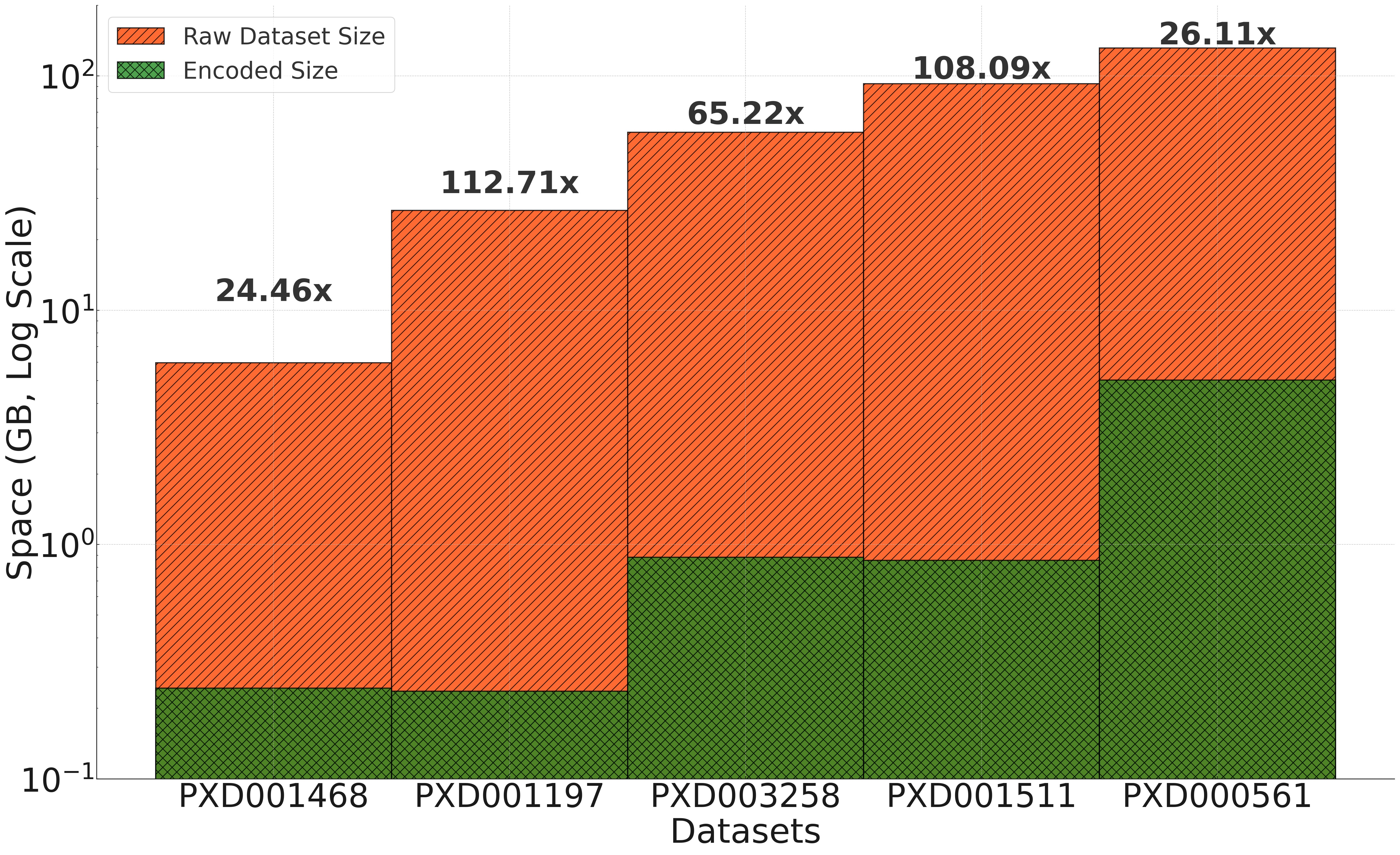}
        \caption{Compression factor}
        \label{fig:compression}
    \end{subfigure}
    \caption{Linkage Efficacy and Compression-SpecHD}
    \label{fig:combined}
\end{figure}

Preliminary tests assessed the best linkage criteria for Hierarchical Agglomerative Clustering (HAC) within our Spec-HD NN-Chain FPGA algorithm. This evaluation stems from the architecture's flexibility to support various linkage criteria, vital for spectral clustering. We fixed an incorrect clustering ratio at 1\% for these tests. Complete linkage proved most effective with a 44\% clustering ratio and 0.764 completeness score. Ward linkage was a close second at 40\% and 0.756, whereas single linkage lagged. For data compression, HV dimensionality (Dhv=2048) was maintained, optimizing resource use, memory, and accuracy. Data compression varied between 24× to 108× across datasets. Within the context of hyperdimensional computing, repeatedly initiating the computational pipeline from the beginning for every analysis proves not only inefficient but also counterproductive. One-time preprocessing and subsequent updates, therefore, emerge as a promising approach for enhancing real-time data analysis.

 \subsection{Speedup comparisons}
In the field of MS-based proteomics, the efficiency of spectral clustering tools is critically measured by runtime, which becomes increasingly important as data repositories continue to expand. With this in mind, we delve into an end-to-end runtime comparison between Spec-HD and several other state-of-the-art tools, specifically targeting GLEAMS for clustering quality and HyperSpec for runtime efficiency, along with msCRUSH and Falcon. Across five datasets, Spec-HD achieves remarkable speed-ups, ranging from 31× over GLEAMS for dataset PXD001511 to an impressive 54× for PXD000561. Against HyperSpec-HAC, the current state-of-the-art in runtime, we note a 6× speed-up, solidifying Spec-HD's efficiency. Our analysis highlights the flexibility advantage of GPU-accelerated encoding, which can be reconfigured during runtime for different tasks. In contrast, FPGA systems like Spec-HD can't be altered on-the-fly due to inherent design constraints. Despite this limitation, Spec-HD offers superior speed, even when constrained to a single encoder module. This speed could be further optimized by utilizing more advanced or multiple FPGAs, increasing its efficiency.

\begin{figure}[htbp]
\centerline{\includegraphics[width=0.45\textwidth]{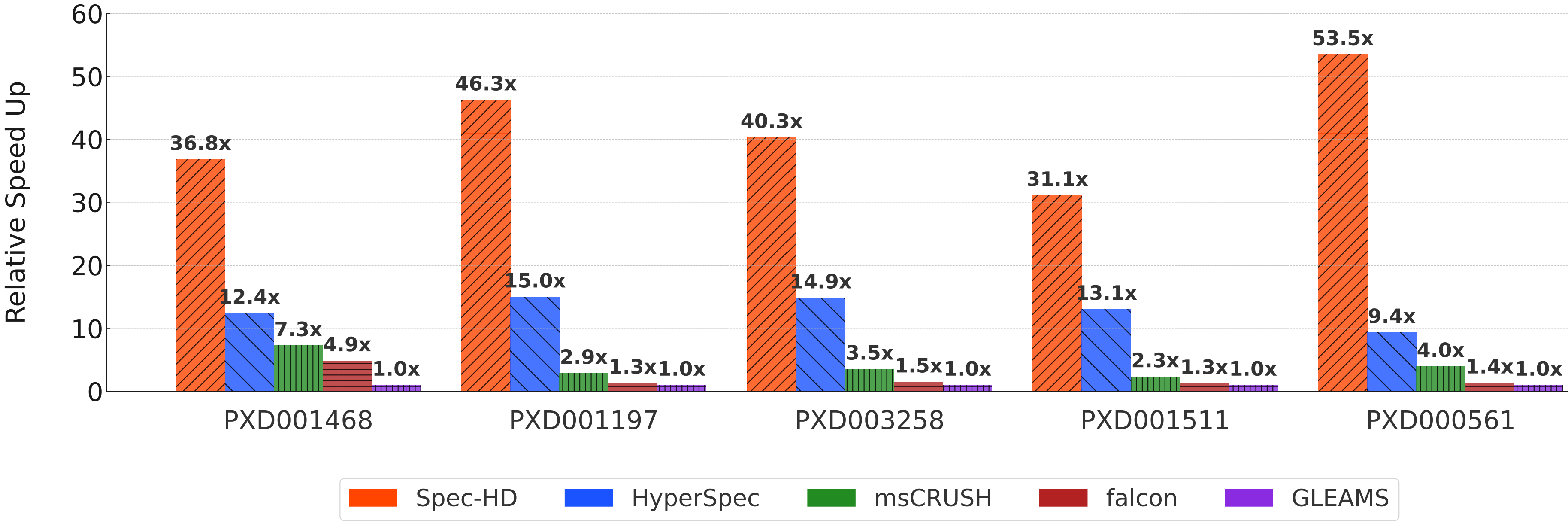}}
\caption{End-to-end runtime speedup}
\label{total_speedup}
\end{figure}

Building on the compelling data compression metrics discussed earlier, our innovative approach suggests that one-time preprocessing can markedly improve downstream analysis efficiency. When concentrating exclusively on standalone clustering of pre-encoded vectors, the runtime gains are remarkable. Spec-HD clocked in at 80 seconds, achieving a 12.3× speed-up in comparison to HyperSpec, which took 1000 seconds. We also note a 14.3× edge over GLEAMS with our largest dataset, PXD00561. These numbers become even more pronounced against Falcon, with 100x speedup, highlighting the framework's prowess in boosting clustering efficiency.

\begin{figure}[htbp]
\centerline{\includegraphics[width=0.35\textwidth]{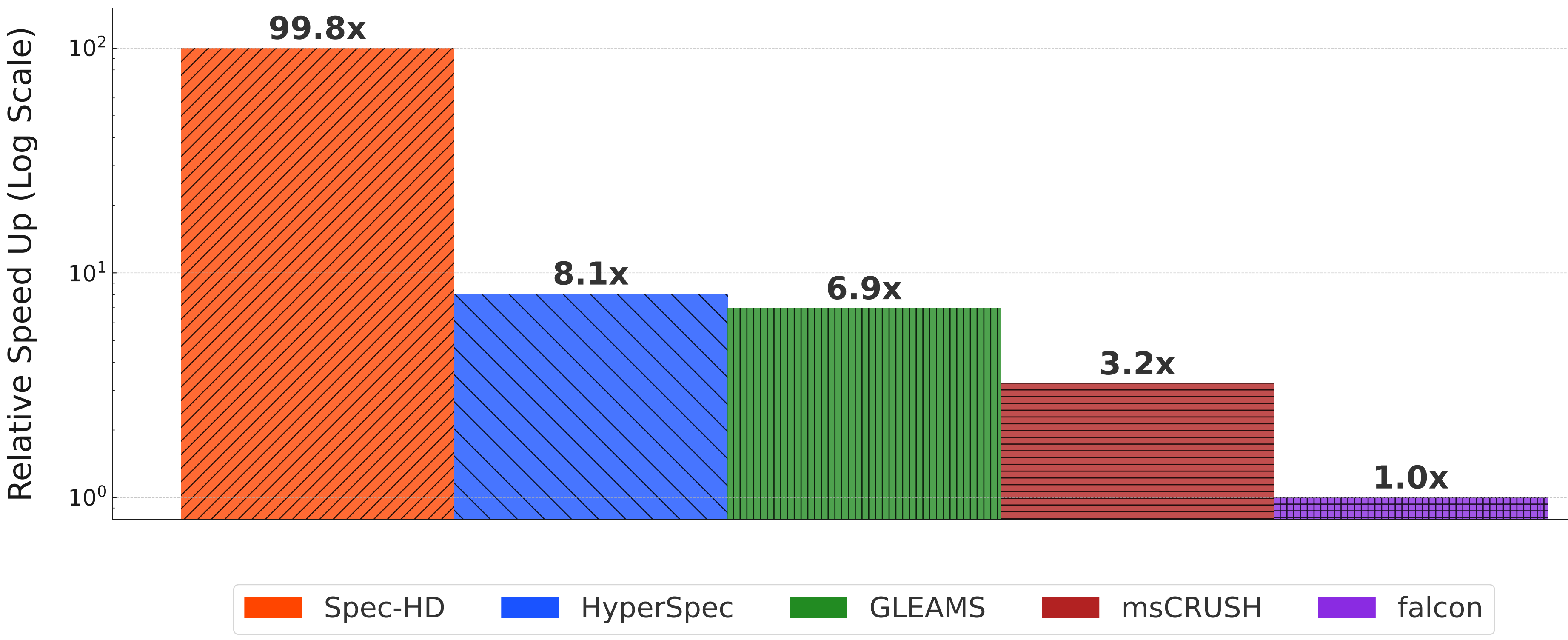}}
\caption{Standalone clustering speedup for PXD000561}
\label{0561}
\end{figure}

\vspace*{-2mm}
 \subsection{Energy efficiency}

In this study, Spec-HD's energy efficiency was evaluated against HyperSpec-HAC and HyperSpec-DBSCAN using Intel RAPL for CPU, Nvidia SMI for GPU, and Xilinx XRT for FPGA measurements. The metrics encompass both end-to-end and clustering-phase efficiencies. Spec-HD exhibited a 14× and 31× improvement in end-to-end energy efficiency over HyperSpec-DBSCAN and HyperSpec-HAC, respectively, with clustering-phase gains of 12× and 40×. These significant improvements are primarily attributed to the efficiency of FPGA configurations. While HyperSpec-DBSCAN demonstrated a threefold lower runtime than HyperSpec-HAC, it lagged in clustering quality, as demonstrated in Fig.~\ref{fig:incorrect_new}. Our results affirm that FPGAs, despite lacking the on-the-fly reconfigurability of GPUs, provide superior energy efficiency—particularly when employing advanced configurations or multiple units. Thus, Spec-HD emerges as a compelling choice for applications where energy efficiency is paramount.
 
 \vspace*{-2.3mm}
\begin{figure}[htbp]
    \centering
    \begin{subfigure}{0.45\linewidth}
        \centering
        \includegraphics[width=0.95\linewidth]{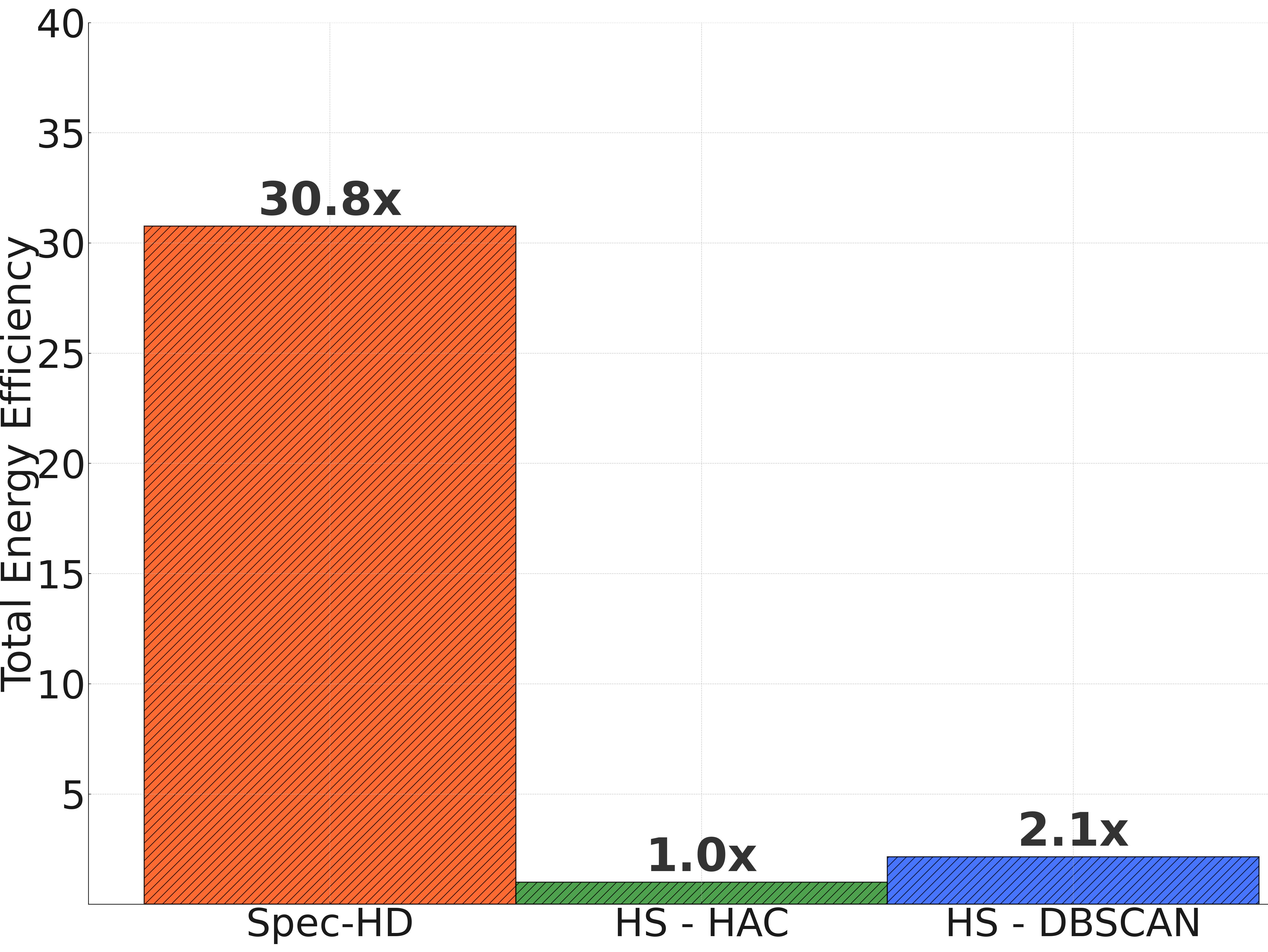}
        \caption{End-to-end}
        \label{fig:totalenergy}
    \end{subfigure}
    \hspace{0.01\linewidth}  
    \begin{subfigure}{0.45\linewidth}
        \centering
        \includegraphics[width=0.95\linewidth]{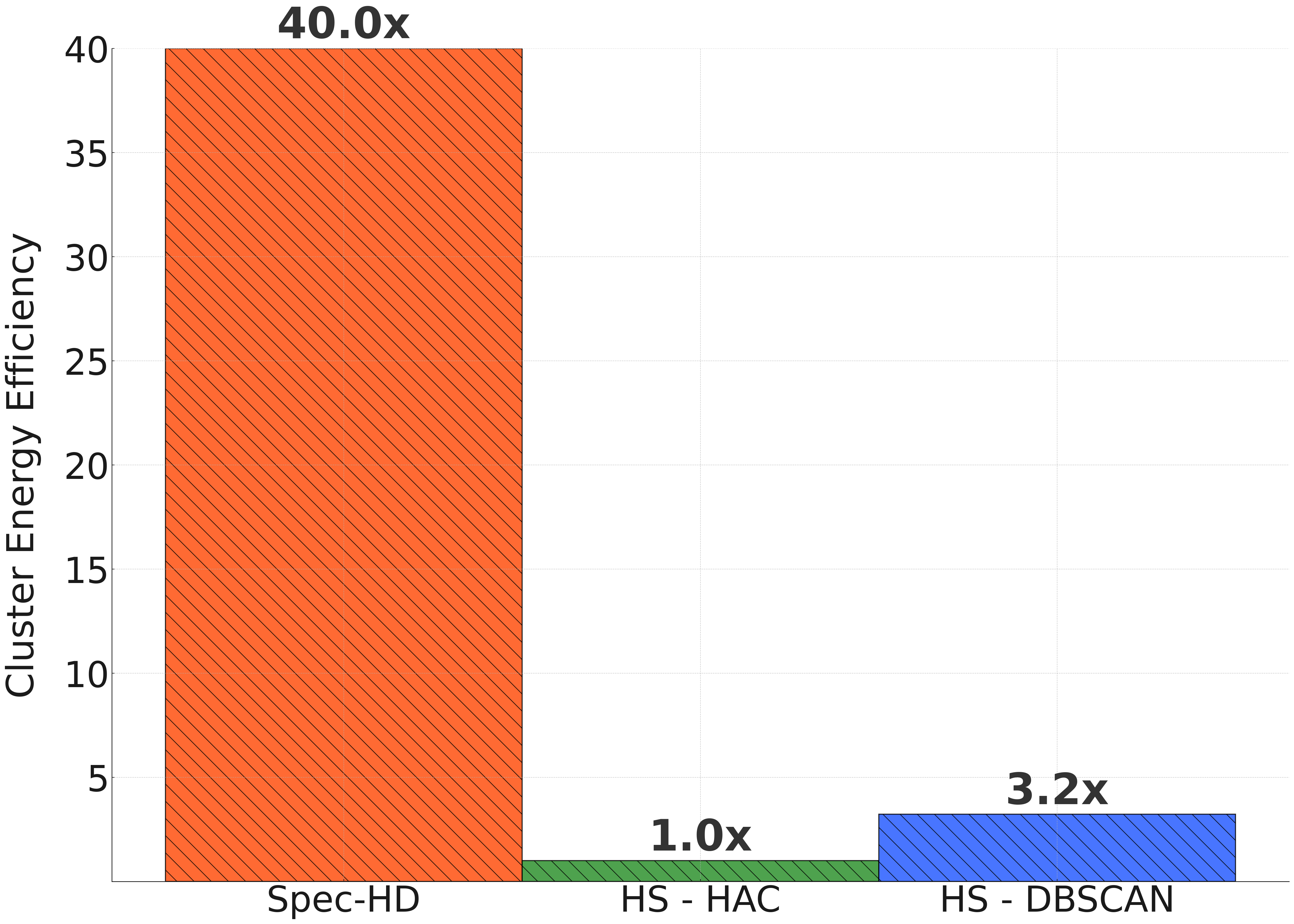}
        \caption{Standalone clustering}
        \label{fig:clusterenergy}
    \end{subfigure}
    \caption{Energy efficiency}
    \label{fig:combinedenergy}
\end{figure}

\vspace*{-2mm}
 \subsection{Clustering quality}
 \subsubsection{Clustered Spectra vs. Incorrect Clustering Ratio}

\begin{figure}[htbp]
\centering
\includegraphics[width=0.34\textwidth]{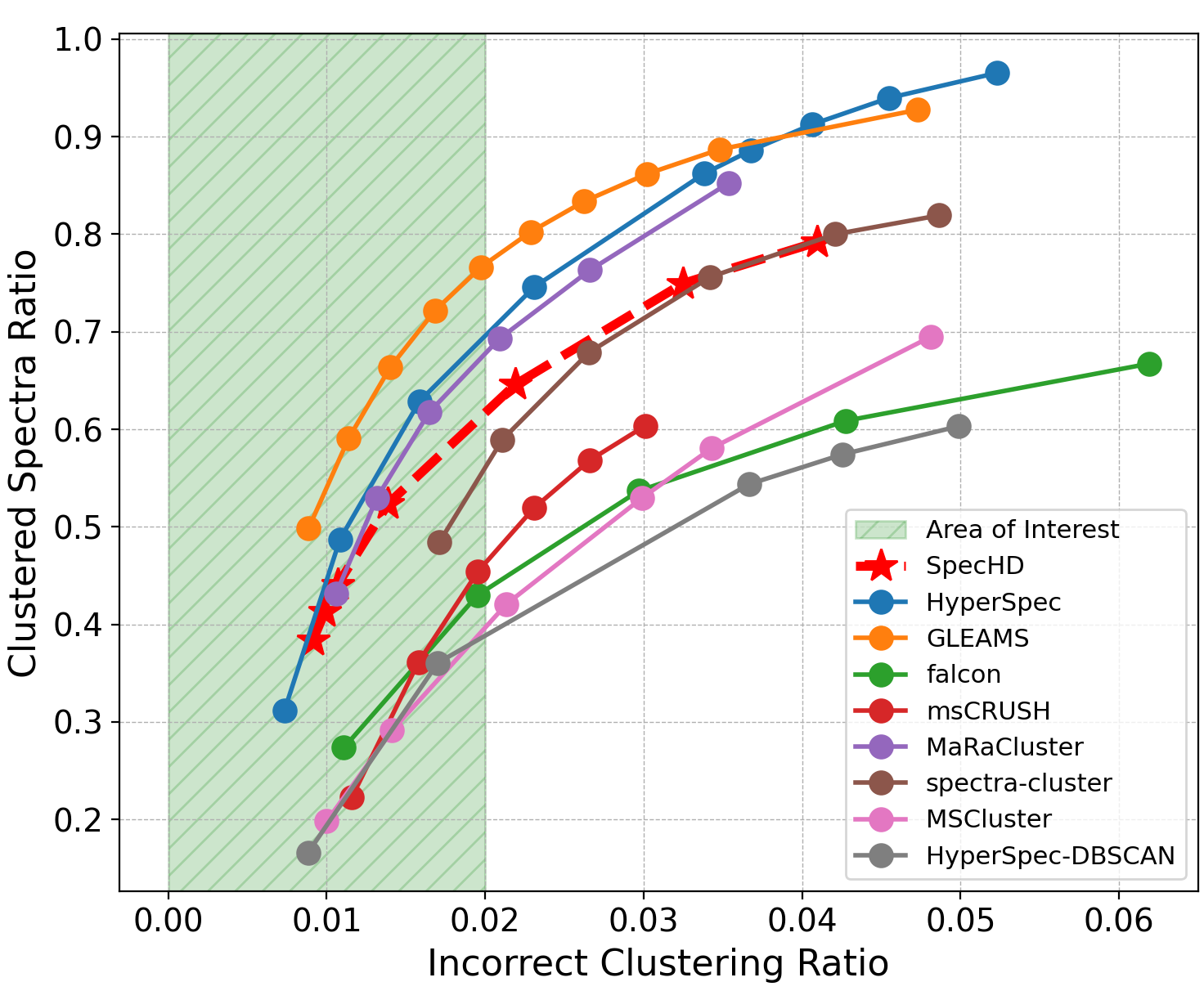}
\caption{Clustered spectra ratio vs incorrect clustering ratio}
\label{fig:incorrect_new}
\end{figure}

Fig.~\ref{fig:incorrect_new} illustrates the balance Spec-HD achieves between clustered spectra ratio and incorrect clustering ratio (ICR). A high clustered spectra ratio, complemented by a low incorrect clustering ratio (typically 1-2\%), signifies the superior capabilities of a clustering algorithm. However, overly aggressive clustering can compromise the analysis quality by increasing incorrect clustering rates. For this analysis, the spectrum identifications were sourced from the MassIVE reanalysis dataset RMSV000000091.3 using the MSGF+ search engine\cite{kim2014msgf}. In a comparison involving nine other tools—HyperSpec-HAC, HyperSpec-DBSCAN, Falcon, msCRUSH, MaRaCluster, spectra-cluster, and MSCluster—we fine-tuned each to operate within an incorrect clustering ratio ranging from 0\% to 7\%.

\vspace*{-2pt}

\vspace*{-1mm}

Aiming for an ICR of approximately 1\% to maintain high fidelity in downstream analyses, Spec-HD achieves a commendable clustered spectra ratio of 45\%. This outperforms several well-regarded tools such as msCRUSH, Falcon, MSCluster, and spectra-cluster. When placed alongside HyperSpec and MaRaCluster, which respectively reach clustered spectra ratios of 48\% and 44\%, Spec-HD holds its ground. While GLEAMS surpasses Spec-HD in clustered spectra ratio, it does so at the cost of extensive time spent on supervised embedding, diminishing its overall efficiency. When viewed in light of Spec-HD's impressive speedup of up to 54x over GLEAMS, this slight compromise in clustering quality becomes far less consequential. The tool achieves a 1.5-2× speedup (ICR = 1-2\%) in spectra searching by skipping redundant searches for similar spectra, underscoring Spec-HD's optimal balance of speed and accuracy in the field.

\subsubsection{Peptide Identification Overlap}

Venn diagrams present an indispensable tool for understanding how effectively Spec-HD, along with other leading tools like HyperSpec and GLEAMS, generate consensus spectra. These consensus spectra serve a vital role in database searches aimed at identifying peptide sequences. For this comparison, we selected the Human Proteome Draft Data set-E as the benchmark. Our analysis reveals that Spec-HD closely trails GLEAMS by a mere 1.38\% for peptides with a precursor charge of 2+ and exceeds HyperSpec's performance by 7.33\% in the same charge category. When focusing on peptides with a precursor charge of 3+, Spec-HD identifies 3.24\% fewer unique peptides compared to GLEAMS but leads HyperSpec by a margin of 5.10\%. These results suggest that Spec-HD effectively balances performance and clustering quality. Upon closer examination of Spec-HD's metrics, we noted its completeness standing at 0.82 for the given parameters, a touch below the typical range of around 0.85 seen in other tools. This nuanced difference not only facilitates the identification of a more diverse array of unique peptides but also reinforces Spec-HD's significant overlap with existing state-of-the-art spectral clustering methods, emphasizing its utility for large-scale human proteome analysis.

\vspace*{-1mm}

 \begin{figure}[htbp]
\centering
\includegraphics[width=0.4\textwidth]{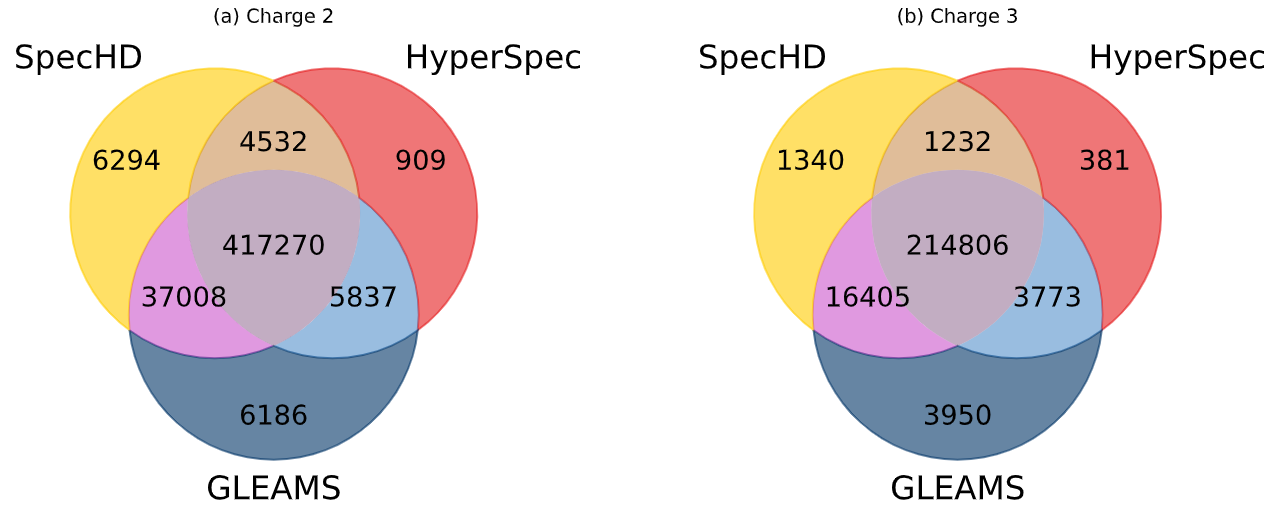}
\caption{Overlap of identified unique peptides}
\label{fig:venn}
\end{figure}

\vspace*{-2mm}
\section{Conclusion}

In summary, this paper introduces Spec-HD, a framework that leverages the power of hyperdimensional computing and FPGA-accelerated architectures to tackle the existing inefficiencies in mass spectrometry-based proteomics. Spec-HD emerges as a high-performing solution, excelling in the critical MS clustering quality metrics and downstream analysis. Our experiments reveal that Spec-HD can efficiently process a vast human proteome dataset in a mere 5 minutes, outperforming other tools while retaining the potential of repository scale clustering. Furthermore, it outclasses current state-of-the-art solutions by offering a speedup factor ranging from 6× to 54× and boasting energy efficiency exceeding 31×. The implications of these advancements are profound, as they present Spec-HD as a catalyst for enabling real-time, high-throughput data analysis that is critical for personalized healthcare applications. Future work should explore further optimization possibilities and integrate Spec-HD at the source of data acquisition within existing proteomics processing pipelines.

\bibliographystyle{IEEEtran}
\bibliography{SpecHD}

\vspace{12pt}

\end{document}